\pdfoutput=1

\documentclass[12pt]{article}
\usepackage[margin=1in]{geometry}

\usepackage{authblk}

\usepackage{amsfonts, amscd, amssymb, amsthm, amsmath, bm, bbm}

\usepackage[colorlinks, citecolor={blue}]{hyperref}

\usepackage[table]{xcolor}
\usepackage{graphicx}

\usepackage[format=plain,
labelfont={bf,it},
textfont=it]{caption}

\usepackage{etoolbox, tikz}
\newtoggle{quickdraw}
\toggletrue{quickdraw} % Uncomment this to render more quickly (non-random)
\usetikzlibrary{matrix, backgrounds, calc, arrows, shapes, decorations, fit, bayesnet, decorations.pathmorphing, decorations.pathreplacing}

\usepackage{booktabs, longtable, multirow}

\usepackage{verbatim, listings}

\usepackage{xr}

\usepackage{natbib}
\usepackage{todonotes}

\usepackage{float}

\title{Synthesizing longitudinal cortical thickness estimates with a flexible and hierarchical multivariate measurement-error model}

\author[1]{Jesse W.~Birchfield}
\author[2]{Nicholas J.~Tustison}
\author[1]{Andrew J.~Holbrook}

\affil[1]{Department of Biostatistics, University of California, Los Angeles}
\affil[2]{Department of Radiology and Medical Imaging, University of Virginia}

\begin{document}

\maketitle
\date{}

\begin{abstract}
MRI-based entorhinal cortex thickness (eCT) measurements predict cognitive decline in Alzheimer's disease with low cost and minimal invasiveness. Two prominent imaging paradigms, FreeSurfer (FS) and Advanced Normalization Tools (ANTs), feature multiple pipelines for extracting region-specific eCT measurements from raw MRI, but the sheer complexity of these pipelines makes it difficult to choose between pipelines, compare results between pipelines, and characterize uncertainty in pipeline estimates. Worse yet, the EC is particularly difficult to image, leading to variations in thickness estimates between pipelines that overwhelm physiological variations predictive of AD. We examine the eCT outputs of seven different pipelines on MRIs from the Alzheimer's Disease Neuroimaging Initiative, with 197 cognitively normal, 324 mildly cognitively impaired, and 142 AD subjects. Because of both theoretical and practical limitations, we have no gold standard by which to evaluate them. Instead, we use a Bayesian hierarchical model to combine the estimates. The resulting posterior distribution yields high-probability idealized eCT values that account for inherent uncertainty through a flexible multivariate error model that supports different constant offsets, standard deviations, tailedness, and correlation structures between pipelines. Among other findings, we discover a tradeoff: ANTs pipeline error distributions have greater standard deviations, while FS pipeline error distributions have heavier tails. Furthermore, our hierarchical model directly relates idealized eCTs to clinical outcomes in a way that propagates eCT estimation uncertainty to clinical estimates while accounting for longitudinal structure in the data. Surprisingly, even though it incorporates greater uncertainty in the predictor and regularization provided by the prior, the combined model reveals a stronger association between eCT and cognitive capacity than do nonhierarchical models based on data from single pipelines alone.
\end{abstract}

\section{Introduction}

Alzheimer’s Disease (AD) afflicts roughly 5.8 million Americans and ranks as the fifth leading cause of death in the US. The CDC projects that, without a cure, this burden will triple by 2060 as the population ages \citep{rodgers2002alzheimer}. Because of the brain’s redundancy (the pattern of distributing memories and abilities among different regions) and plasticity (the ability for one region to compensate for damage to another), it takes fairly extensive brain damage to cause obvious cognitive impairment. Thus, AD patients tend not to notice symptoms and only seek diagnosis several years into the disease \citep{rodgers2002alzheimer}.  Existing therapies can slow the course of the disease and delay the appearance of symptoms, but the effectiveness of these therapies hinges on early diagnosis \citep{rodgers2002alzheimer}.

During the presymptomatic phase of AD, beta-amyloid plaques and neurofibrillary tangles begin destroying neurons, and the cerebrum begins to atrophy \citep{sahyouni2017alzheimer}. The damage begins locally and slowly spreads. Neuroscientists once identified the hippocampus as the starting point of this process. They have since discovered that it actually begins in the entorhinal cortex (EC), a gateway structure between the hippocampus and the neocortex \citep{rodgers2002alzheimer}. \citet{holbrook2020anterolateral} offer evidence that pinpoints certain EC substructures  as the site of the earliest AD lesion.

Imaging of the EC is a vital, noninvasive, and inexpensive tool for early AD detection. For static representations of anatomical structures, MRI provides the best resolution of all imaging technologies. Yet a phenomenon as subtle as incipient AD damage is difficult to distinguish. The difference in EC thickness between a healthy person and an AD patient is on the order of 1 to 2 millimeters \citep{holbrook2020anterolateral}. Based on the same MRI, different cortical thickness (eCT) estimation techniques give results with a range greater than this \citep{holbrook2020anterolateral}.

More precise eCT estimation, then, could directly enable earlier AD detection. As MRI hardware improves, so must the software that transforms sets of electromagnetic signals into three-dimensional images (registration), with tissue types differentiated (segmentation), with regions labeled and measured. Two open-source toolkits or “ecosystems” have become standard for this task: FreeSurfer (FS) and Advanced Normalization Tools (ANTs). Within each ecosystem, developers experiment with different ways of leveraging data—such as creating a template from a population, or from repeated measures of the same subject—and different computational methods—such as  networks—to measure eCT. We call these discrete methods “pipelines.”

\citet{tustison2019longitudinal} select seven pipelines to process 2515 raw MR images from the ADNI-1 study and to compute a vector of 2515 EC thickness measurements. They devise metrics and statistical tests to evaluate different pipelines’ outputs on the criteria of (i) reliability, or consistency of measurements on the same subject, and (ii) performance when used in a statistical model as a predictor of a known quantity. The problem, as they emphasize, is that neither of these is the same as accuracy. We cannot compare these numbers to the real cortical thicknesses. We are in a situation with no “ground truth.”

In this paper, we take a different strategy. Rather than judging pipeline quality, we assume \emph{a priori} that each individual pipeline offers its own important information on eCT. The challenge then becomes pooling the respective pipeline contributions together in a principled manner that enables quantification of the uncertainty inherent in eCT estimation. We use the seven vectors of EC measurements as a case study. Given our desiderata, a good tool for estimating eCT from our seven sets of estimates is the Bayesian hierarchical model. In Bayesian inference, prior probability distributions represent our beliefs about plausible parameter values before seeing the data. Posterior distributions represent how a rational agent would update his prior beliefs upon seeing the data. In a hierarchical model, each group (here, pipeline) has an error distribution, indexed by its own parameter. These parameters themselves are draws from an underlying distribution, indexed by a hyperparameter. We estimate all the quantities simultaneously, so each observation influences the estimation of its own group parameter, and through it the hyperparameter, and indirectly the other group parameters. In this way, every data point informs the estimation of every parameter. In a well-specified model, using all of the information allows for more accurate and precise results. Putting these features together, the Bayesian hierarchical model allows us to encode our assumptions about parameter values as priors, and our assumptions about parameter relationships as structural choices. Although the particular assumptions we make in our model specification are open to challenge—indeed, we hope they will be challenged and refined—we believe that the Bayesian hierarchical model is an effective approach in principle, one that advances the science of eCT estimation insofar as it relates to clinical outcomes.

Cortical thickness has scientific importance in its own right, but its clinical importance in the context of AD research is in how well it predicts AD symptoms. One measure of these symptoms is the patient’s score on the Mini Mental State Examination (MMSE) \citep{tombaugh1992mini}. Given the demonstrated imprecision of eCT data, to what extent can we still use it for inferences about the relationship between eCT and other variables, such as MMSE? Again, the Bayesian hierarchical model provides a way forward. Idealized eCT becomes a latent parameter; in other words, our posterior distribution for idealized eCT, estimated from the seven pipelines, becomes a covariate at a “higher” level of the hierarchical model. This is a longitudinal model that predicts change in MMSE over time as a function of eCT, while controlling for age, sex, diagnostic category, and random effects (individual differences). By propagating the uncertainty in eCT into the uncertainty of the coefficient estimates, the Bayesian framework helps us avoid overconfidence and spurious conclusions.

\section{Data and data processing}

\subsection{ADNI-1 and MMSE}

Our data comes from the Alzheimer's Disease Neuroimaging Initiative \citep{jack2008alzheimer, petersen2010alzheimer, mueller2005ways}, a public-private partnership sponsored in part by the National Institute on Aging (NIA). Beginning in October 2004, the three-year longitudinal ADNI-1 study followed a cohort of North American patients comprising 200 elderly controls, 400 with mild cognitive impairment (MCI), and 200 with diagnosed Alzheimer's Disease (AD). At each visit, researchers conducted cognitive assessments and structural MRI scans. Our analysis uses data from 663 of the 800 patients tabulated and made publicly available in \citet{tustison2019longitudinal}, with a total of 2449 visits. We simply omit those visits that had missing MMSE scores.

The MMSE is a screening test for cognitive impairment. It has 10 questions, takes 5 to 10 minutes to administer, and covers space and time orientation, naming familiar objects, repeating back a phrase, recalling a previous question, manipulating letters and/or numbers, and following instructions. Out of 30 possible points, a score of 24 or more indicates normal cognition, 19-23 indicates mild cognitive impairment, and 18 or below indicates moderate or severe impairment. While it can help differentiate different types of dementia, the test itself is not sufficient for diagnosis.

\subsection{Description of seven pipelines} \label{sec:pipeline_descriptions}

We analyze output from seven eCT pipelines. FreeSurfer and ANTs use different geometrical constructs to stand in for eCT. FreeSurfer is mesh based, using polygonal meshes representing the gray/white matter surfaces and outer cortical surfaces \citep{fischl2012freesurfer}. We denote the FreeSurfer pipelines as FSCross and FSLong, where ``cross" means ``cross-sectional" and ``long" means ``longitudinal." In general, a cross-sectional pipeline interprets each image independently, not taking into account repeated measures, whereas a longitudinal pipeline incorporates additional considerations, e.g. creating a single-subject template (SST) for all images of the same subject, to reduce within-subject variability.

The pipelines of the ANTs ecosystem are volumetric, using diffeomorphic mappings \citep{das2009registration}. Like FSCross, ANTsCross use a cross-sectional pipeline that treats images from the same individual at different time points as independent. ANTsSST adds a longitudinal component including ``rigidly transforming each subject to the SST and then segmenting and estimating cortical thickness in the space of the SST" \citep{tustison2019longitudinal}. ANTsNative adds longitudinal weighting without an SST to ``segment and then estimate cortical thickness in the native space" \citep{tustison2019longitudinal}. ANTsNetCross follows a similar algorithm to ANTsCross, but for computation uses an artificial neural network to replace some steps, e.g. template-based brain extraction, template-based n-tissue segmentation, and joint label fusion. Likewise, ANTsXNetLong brings the neural net to ANTsSST. The X stands for R or Python, as it can interface with either programming environment.

\subsection{Exploration of pipeline outputs}

\begin{figure}[!ht]
	\centering
	\includegraphics[width=0.7\linewidth]{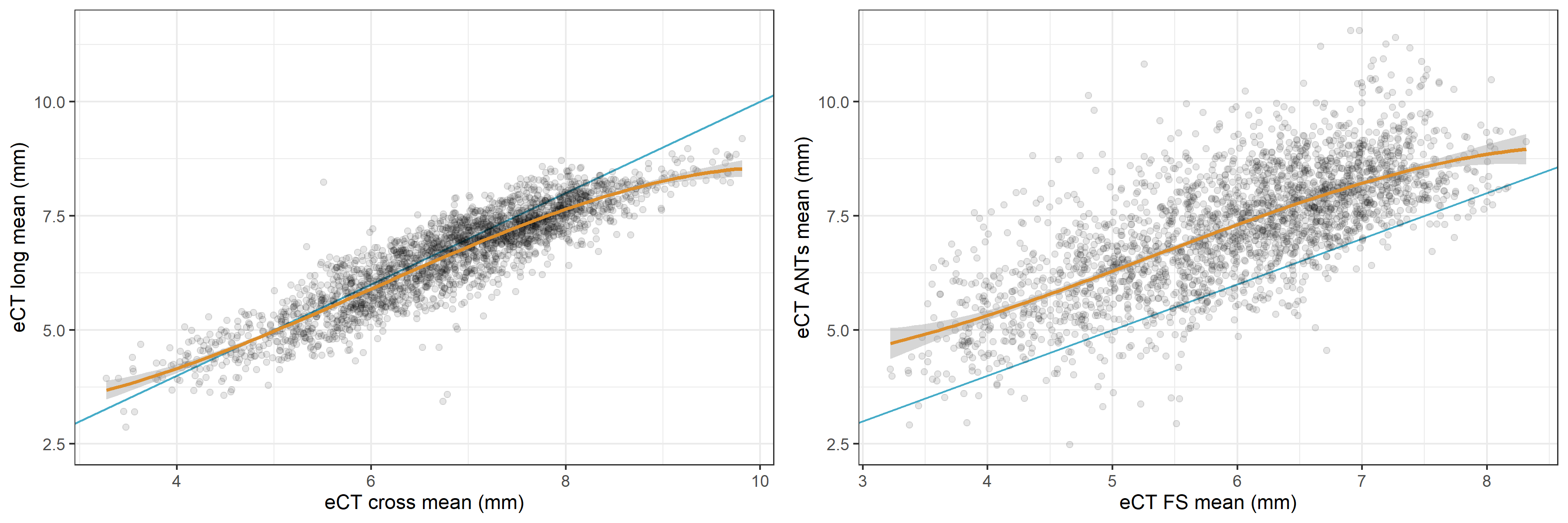}

	\caption{With scatterplots and smoothed curves, two categorizations of the pipelines, to compare average cortical thickness estimates by category. The left plot shows that longitudinal pipelines give slightly higher estimates at higher levels of cortical thickness. The right plot shows that Advanced Normalization Tools (ANTs) pipelines give much higher estimates at all levels of cortical thickness (eCT) than do FreeSurfer (FS) pipelines.}
	\label{fig:pipeline_comparison_scatter}
\end{figure}

\begin{figure}[ht]
	\centering
	\includegraphics[width=0.7\linewidth]{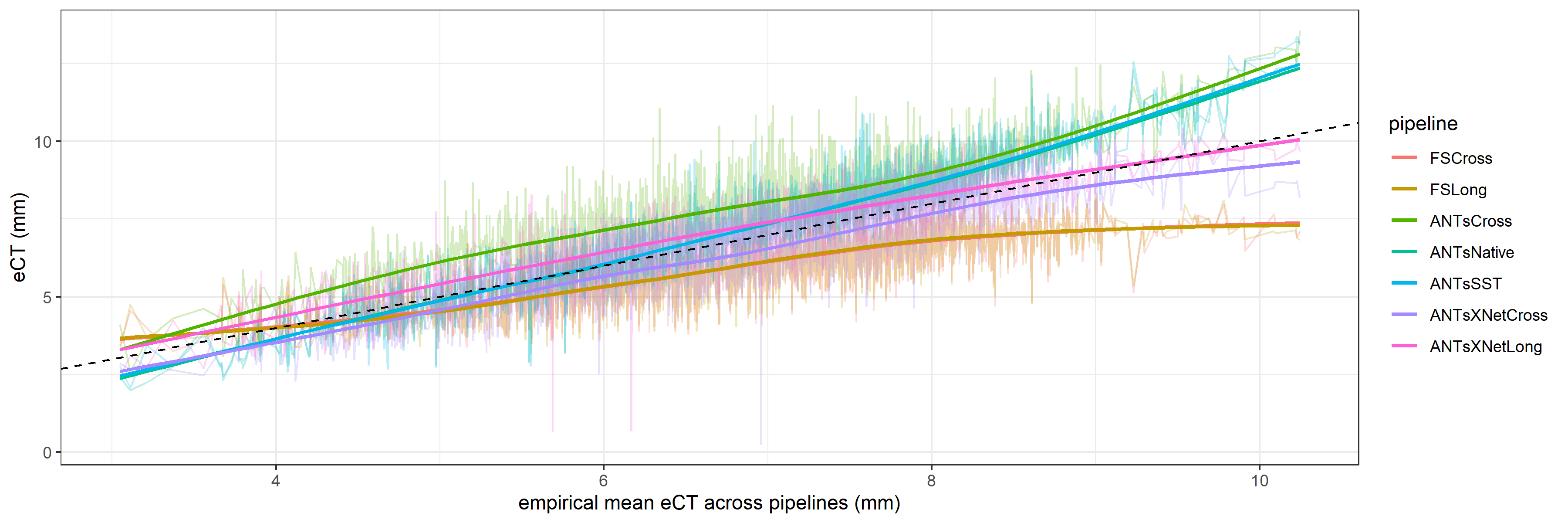}

	\caption{Observations arranged by their empirical mean and raw pipeline measurements as jagged lines, overlaid with local averages as locally estimated scatterplot smoothed (LOESS) curves. Pipelines are variants of FreeSurfer (FS) and Advanced Normalization Tools (ANTs). Across the bulk of the data, there is evidence of (i) roughly constant offsets, (ii) roughly constant error variance across cortical thickness (eCT) values within each pipeline, (iii) some outliers, (iv) low correlation of errors within pipelines, and (v) high correlation of errors across pipelines. These observations are consistent with our modeling assumptions.}
	\label{fig:prior_spaghetti}
\end{figure}

\begin{figure}[ht]
	\centering
	\includegraphics[width=0.7\linewidth]{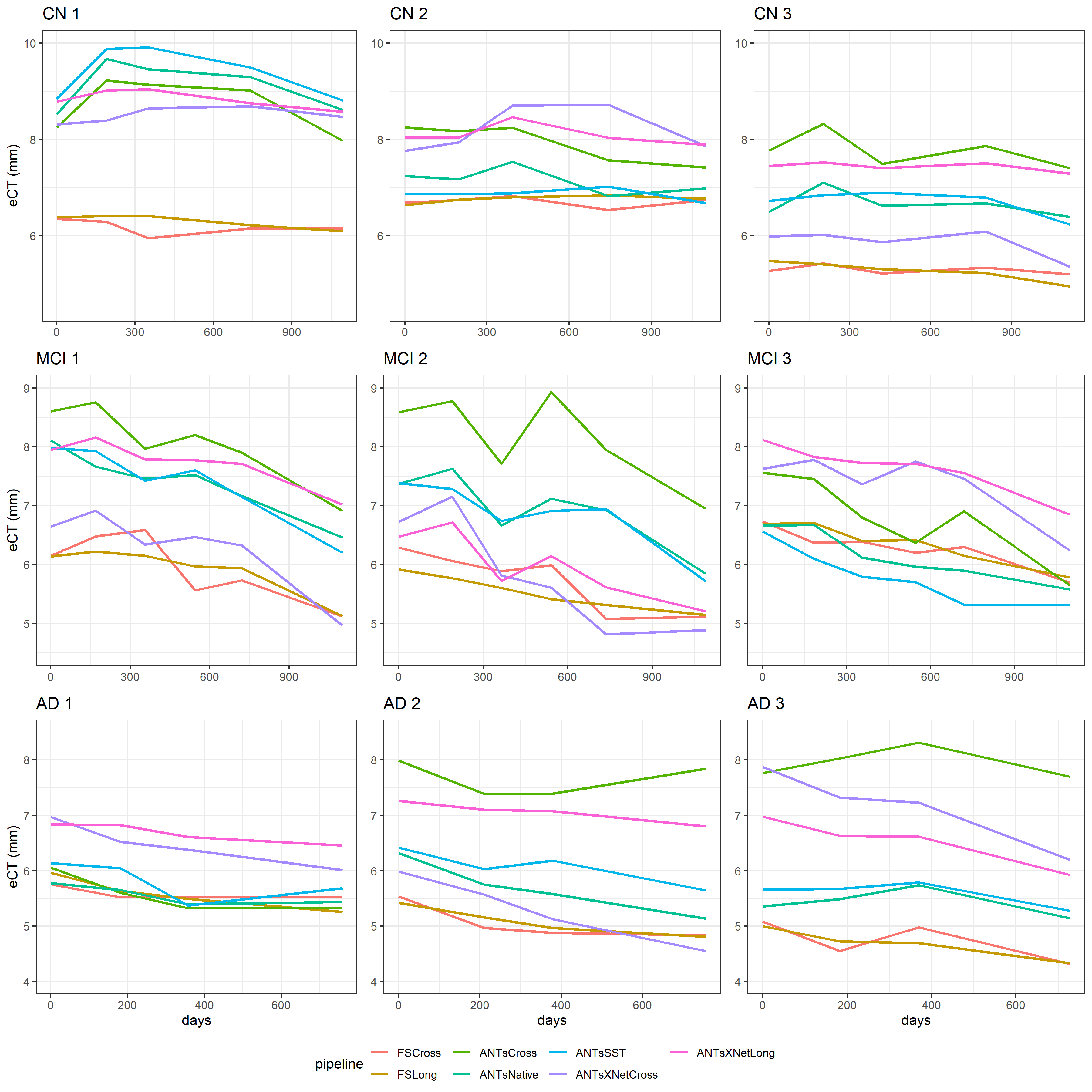}

	\caption{Longitudinal profile plots of cortical thickness (eCT) for nine randomly selected individuals, three from each diagnostic category: cognitively normal (CN), mildly cognitively impaired (MCI), and Alzheimer's disease (AD). Pipelines are variants of FreeSurfer (FS) and Advanced Normalization Tools (ANTs). The profiles run roughly in parallel, often in similar sequence, but with correlated errors.}
	\label{fig:prior_profiles}
\end{figure}

We structure our data as follows. Each observation, or row, consists of 7 different eCT estimates computed from a single scan from a single subject at a single time point. The value estimated is the sum of the left EC thickness and the right EC thickness, in millimeters. Each column consists of the 2449 estimates from a single pipeline.

The seven pipelines can be classified in different ways, including cross-sectional vs. longitudinal and FreeSurfer vs. ANTs. In Figure \ref{fig:pipeline_comparison_scatter}, each point on a scatterplot represents a single observation. The blue lines represent identity, and the orange lines are LOESS curves \citep{cleveland1988locally}. Of the seven different eCT measurements, the left plot compares the mean of the longitudinal pipelines to the mean of the cross-sectional pipelines. On average, cross-sectional pipelines tend to give lower estimates at higher levels of eCT. The right plot compares the mean of the FreeSurfer pipelines to the ANTs pipelines. ANTs pipelines, on average, tend to give higher eCT estimates for every level of eCT.

To create Figure \ref{fig:prior_spaghetti}, we arrange the rows by their empirical mean, plot the estimates vertically, and connect them with horizontal lineplots, the thinner lines. The mean (not shown) would be the straight line $y=x$. We also overlay the lineplots with LOESS curves with a high degree of smoothing. These thicker lines convey the average pipeline-specific estimates over a small interval of mean eCT.

Comparing the LOESS curves to each other, we observe some differences between the pipelines. The curves for FSCross and FSLong are nearly indistinguishable, as are the curves for ANTsNative and ANTsSST. When the mean is above 5mm (true of 93\% of the observations), FSCross and FSLong give the lowest estimates. Except for the very low end, where the curve is based on only a handful of points, ANTsCross gives the highest estimates. ANTsNative and ANTsSST have the greatest range, from about 2.5 to about 12.5, and FSCross and FSLong have the least range, from about 3.75 to about 7.5. Most relevant for our statistical modeling, the middle 90\% of mean eCT measurements fall in the interval (4.74, 8.61). On this interval, the curves run roughly in parallel, with the exception of ANTsXNetLong. The pipelines have a roughly constant offset from each other. On the reasonable assumption that a plot of the idealized eCT values would also run parallel, we can say that each pipeline has a roughly constant offset from the ideal.

Comparing the line plots to the LOESS curves, especially in the region that includes most of the data, we notice first that they are jagged. Adjacent observations for the same pipeline are frequently 2mm or more apart and on opposite sides of the curve. This is evidence that the within-pipeline deviations from the local average are uncorrelated as a function of the empirical mean, or perhaps only weakly correlated. Were they strongly correlated, the line plots would stay on one side of the local average for longer. The majority of the deviations stay uniformly within a certain distance, perhaps 1mm, of the local average, regardless of its magnitude. This supports our choice (Section \ref{sec:ct_model}) to model errors arising from the same pipeline as identically distributed. Because of the presence of outliers, visible as long spikes on the line plots, we choose a family of probability distributions with heavier tails that makes outliers less improbable than does the normal (more on this below). Across pipelines, however, we see strong correlations. In many places, certain line plots tend to zig and zag together. FSCross and FSLong show this effect clearly, as do ANTsNative and ANTsSST. We should expect it, since the different pipelines share assumptions and techniques to varying degrees. But if we fail to account for these correlations in our model and erroneously treat the pipelines as independent, we will give inordinate weight to certain measurements just for being duplicated.

We can see different patterns if we separate observations by individuals and order them by time. Figure \ref{fig:prior_profiles} shows profile plots of eCT over time for three individuals randomly selected from each of the diagnostic categories (CN, MCI, AD). On a given plot, each line represents the sequence of eCT measurements from a given pipeline. The profiles run roughly in parallel, which supports our assumption of constant pipeline offsets. They also contain some jaggedness and even local upward motion, which is much more likely to represent measurement error than an actual gain of cortical thickness in the subject. This supports our assumption of random errors, uncorrelated within pipelines but correlated between pipelines. With these considerations in mind, we develop a multivariate measurement error model that accounts for offsets, correlations across pipelines, and differential tailedness between pipelines.

\section{Hierarchical model development}

\subsection{Robust error estimation} \label{sec:tdist}

Typically, in regression models, random errors have the Gaussian or normal distribution. But when the data contain outliers, there is a more robust option for modeling errors. Consider that the density of the standard normal distribution at 0 is about .40, and at 3 it is about .004. A $z$-score of 3 is roughly 100 times less probable than a $z$-score of 0. If we use a normal distribution, we are encoding a belief that observations more than a couple of standard deviations from the mean are extremely rare. In such a model, outliers tend to pull the estimated mean away from the center of the data and inflate the estimated variance \citep{gelman1995bayesian}.

While the normal distribution has only two parameters, the location-scale Student-$t$ distribution has three: mean, scale ($\tau$), and degrees of freedom ($\nu$). If $\nu$ is small, the curve has heavy tails. For example, if $\nu=1$, then a $z$-score of 3 is only 10, not 100, times less likely than a $z$-score of 0. On the other hand, if $\nu$ is large, say 30 or more, then the curve is virtually identical to the normal with the same mean and scale. In fact, the $t$ is a generalization of the normal, because it converges in distribution to the normal as $\nu\rightarrow\infty$. So using a $t$ likelihood with $\nu$ to be estimated reflects open-mindedness about the frequency of outliers. It allows for the distribution to be normal, or not. And when it is not, the $t$ gives an estimate of the mean that is nearer the center of the data \citep{kruschke2014doing}.

When errors may be correlated, we extend the normal model to the multivariate normal model, which includes a covariance matrix $ \bm{\Sigma} $ as a set of parameters to be estimated. Suppose $ \bm{Y} $ is a vector of $K$ observations. It can be shown that $ \bm{y} \sim N_{K}(\bm{\mu}, \bm{\Sigma}) $ is equivalent to the scale mixture form $ \bm{y} = \bm{\mu} + \bm{\Sigma}^{1/2}\bm{z} $, where $ \bm{z} \sim N_{K}(\bm{0}, \bm{I}) $ is a vector of i.i.d. standard normal random variables. Furthermore, we can factorize the covariance matrix as $ \bm{\Sigma} = \bm{T R T} $, where $\bm{T}$ is the diagonal matrix with standard deviations $\tau_1, ..., \tau_k$ on the diagonal, and $\bm{R}$ is the correlation matrix of $\bm{y}$ \citep{barnard2000modeling}. Since the correlation matrix is symmetric with 1's on the diagonal, there are $ \binom{K-1}{2} $ correlation parameters to estimate, indexed as $ \rho_{k k'} $ with $k>k'$ for $k=1,...,K$ and $k'=1,...,K-1$ .

The equivalent extension of the location-scale $t$ model is the multivariate $t$ model, $\bm{y} \sim t_{k}(\nu, \bm{\mu}, \bm{\Sigma})$, which has scale mixture form $\bm{y} = \bm{\mu} + q^{-1/2}\bm{\Sigma}^{1/2}\bm{z}$, where $q \sim \chi^{2}_{\nu} / \nu$ and $ q \perp \bm{z}$. Parameters $\bm{\mu}$ and $ \bm{Z} $ and $ \bm{\Sigma}$ have the same meaning, and the latter has the same factorization. Although it is more flexible than the multivariate normal, it is not quite flexible enough for our purposes. It constrains $\nu$ to be the same for every element of $y$. Since we observe differential tailedness (Figure \ref{fig:prior_spaghetti}), we want to estimate a distinct $\nu_{k}$ for each pipeline. Recent developments \citep{jiang2016robust} show that a certain straightforward generalization of the scale mixture form of the multivariate $t$ has the  properties we need. It has different degrees of freedom in different dimensions, and each dimension is marginally a location-scale $t$ distribution. Instead of a single parameter $ q \sim \chi^{2}_{\nu} / \nu $, let $ q_{k} \sim \chi^{2}_{\nu_{k}} / \nu_{k} $ for $k = 1,...,K$. Let $\bm{Q}$ be the diagonal matrix with these on the diagonal. Then define the ``non-elliptically-contoured $t$ distribution" (NECT) by its scale mixture form: $ \bm{y} \sim NECT_{k}(\bm{\nu}, \bm{\mu}, \bm{\Sigma}) $, with vector $\bm{\nu}$, is defined to be equivalent to $\bm{y} = \bm{\mu} +  \bm {Q}^{-1/2} \bm{\Sigma}^{1/2} \bm{z}$. We now have the tools to specify the full hierarchical model.

\subsection{A model for idealized CT and measurement errors} \label{sec:ct_model}

We model the relationship between estimated eCT and idealized eCT as follows:
\begin{center}
	estimated eCT = idealized eCT + pipeline offset + NECT random error.\\
\end{center}
Let $i$ index the subjects $(i=1:663)$ and $j$ index the visits, for 2449 $(i,j)$ pairs. Let $k$ index the pipelines $(k=1:7)$. Denote the (unobserved) idealized eCTs as $eCT_{ij}$ and the observed eCTs as $\widehat{eCT}_{ijk}$. Each pipeline has its own offset, $\phi_k$. Marginally, each pipeline's errors have an i.i.d $t$ distribution, with location parameter $ \phi_{k} $, scale parameter $\tau_k$, and degrees of freedom parameter $\nu_k$. Thus we write
\begin{gather*}
	\widehat{eCT}_{ijk}=eCT_{ij}+\delta_{ijk} \\
	\delta_{ijk} \sim t(\phi_k, \tau_k, \nu_k)
\end{gather*}
To be clear, we have $7\times2449=17143$ observations in $\widehat{eCT}$. We use them to estimate 2449 latent variables in $\bm{eCT}$, and 7 parameters each in $\bm{\phi}$, $\bm{\tau}$, and $\bm{\nu}$.

To model the correlation between pipelines, let $\widehat{\textbf{eCT}}_{ij*}$ be one of the 2449 rows of the data matrix, transposed to a column vector: the seven eCT estimates computed from a single scan from a single subject at a single time point. Let \textbf{1} be the vector of ones, let $\bm{\phi}$ be the vector of offsets, and let $\bm{\Sigma}$ be a $7\times7$ covariance matrix. The model is
\begin{gather*}
	\widehat{\textbf{eCT}}_{ij*} \sim NECT_{7} (\bm{\nu}, eCT_{ij} \textbf{1} + \bm{\phi}, \bm{\Sigma}).
\end{gather*}
We factor $ \bm{\Sigma} $ as described above, and we choose weakly informative and identical priors for each pipeline, plus a regularizing prior for $\bm{eCT}$ itself:
\begin{gather*}
	\phi_k \sim N(0,3), \quad \tau_k \sim N^{+}(0,3), \quad \nu_k \sim Expo(\textrm{mean=30}), \quad k=1:7 \\ \bm{R} \sim LKJCorr(2), \quad eCT \sim N(7,2).
\end{gather*}
As a prior for $\bm{R}$, we use the Lewandowski-Kurowicka-Joe distribution, a single-parameter family of positive definite matrices \citep{gelman1995bayesian}. Our priors conservatively communicate that (i) the offsets are most likely (i.e. with about 95\% probability) between -6 and 6, and more likely to be nearer 0 than farther from it, (ii) the scale parameters are most likely less than 6, and more likely to be nearer 0 than farther from it, and (iii) the error distributions are about equally likely to be normal as heavy-tailed, and (iv) the errors are somewhat more likely to be positively correlated than independent or negatively correlated.

\tikzstyle{latent}+=[color=blue!75,fill=blue!20,text=black,thick]
\tikzstyle{obs}+=[color=red!75,fill=red!20,text=black,thick]
\tikzstyle{det}+=[color=black!75,fill=black!20,text=black,thick]
\tikzstyle{plate}+=[color=black!75,text=black,thick]
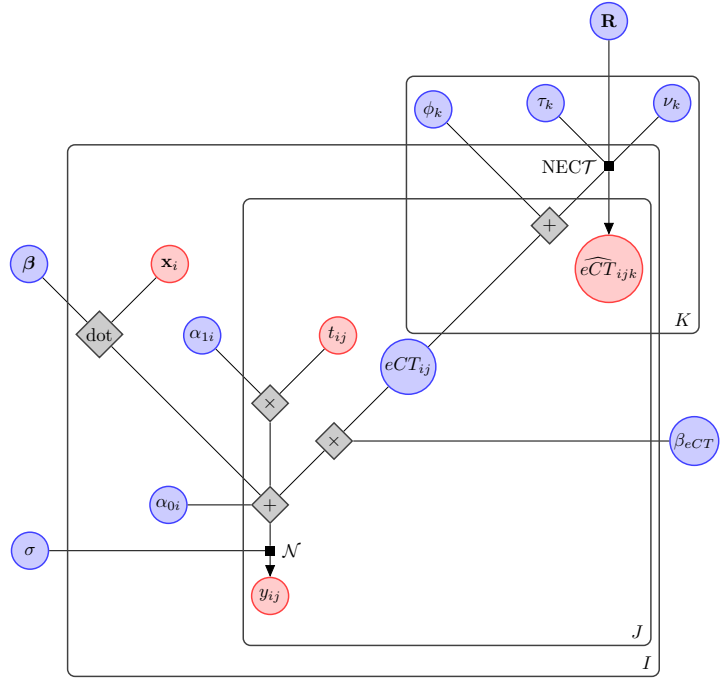
\begin{figure}[t]
	\centering
	\scalebox{0.7}{	\begin{tikzpicture}
			% Y
			\node[obs] (y)   {$y_{ij}$}; %
			\factor[above=of y] {y-f} {right:$\mathcal{N}$} {} {} ; %
			% W and X
			\node[det, above=of y] (plus2) {$+$} ; %
			\node[latent, left=4.1 of y-f] (s) {$\sigma$}; %
			\node[det, above right=1.2 of plus2] (times1) {$\times$} ; %
			\node[latent, above right=1.2 of times1] (ct) {$eCT_{ij}$}; %
			\node[latent, right = 6 of times1] (betaCt) {$\beta_{eCT}$}; %
			\node[det, above=1.2 of plus2] (times2) {$\times$} ; %
			\node[obs, above right=1.2 of times2] (timeIJ) {$t_{ij}$} ; %
			\node[latent, above left=1.2 of times2] (alpha1) {$\alpha_{1i}$} ; %
			\factoredge {plus2,s} {y-f} {y} ; %
			\node[det, above right=3 of ct]            (plus) {$+$} ; %
			\factor[above right=1.2 of plus] {ctHat-f} {left:NEC$\mathcal{T}$} {} {} ; %
			\node[obs, below=1.2 of ctHat-f]    (ctHat) {$\widehat{eCT}_{ijk}$}; %
			\node[latent, above left=2.5 of plus]    (offset) {$\phi_{k}$}; %
			\edge[-] {ct,offset} {plus};
			\edge[->] {ctHat-f} {ctHat};
			\node[latent, above left=1.2 of ctHat-f]    (tau) {$\tau_{k}$}; %
			\node[latent, above right=1.2 of ctHat-f]    (nu) {$\nu_{k}$}; %
			\node[latent, above=2.3 of ctHat-f]    (R) {$\mathbf{R}$}; %
			\edge[-] {plus,nu,tau,R} {ctHat-f};
			\node[latent, left=1.2 of plus2]    (alpha0) {$\alpha_{0i}$}; %
			\node[det, above left=4 of plus2] (dot) {dot};
			\node[obs, above right=1.2 of dot] (x) {$\mathbf{x}_i$};
			\node[latent, above left=1.2 of dot] (beta)   {$\boldsymbol{\beta}$}; %
			\edge[-] {ct} {times1} ;
			\edge[-] {plus2} {times1} ;
			\edge[-] {times1} {betaCt} ;
			\edge[-] {times2} {plus2} ;
			\edge[-] {times2} {timeIJ} ;
			\edge[-] {times2} {alpha1} ;
			\edge[-] {plus2} {alpha0} ;
			\edge[-] {dot} {beta} ;
			\edge[-] {dot} {x} ;
			\edge[-] {dot} {plus2} ;
			% Plates
			\plate {J} {
				(y) %
				(ctHat) %
				(ct) %
				(plus)
			} {$J$} ;
			\plate {I} { %
				(y)(y-f)(y-f-caption) %
				(ctHat)(ctHat-f)(ctHat-f-caption) %
				(dot) %
				(J)
				(alpha0)
				(x)
			} {$I$} ;
			\plate {K} {%
				(ctHat) %
				(ctHat-f)
				(nu) %
				(tau) %
				(offset)
			} {$K$} ;
	\end{tikzpicture}}
	
	\caption{Bayesian plate diagram of the core  model. A node on a plate denotes a repeated variable: $i$ indexes subjects, $j$ indexes time points, and $k$ indexes pipelines. A red node denotes an observed quantity, and a blue node an unobserved quantity. The node $\bm{x_i}$ denotes the vector of observed covariates: age, sex, diagnostic category, time, and the interactions of diagnostic category with time. eCT is cortical thickness, and NECT is the non-elliptically-contoured $t$ distribution explained in Section \ref{sec:tdist}.}
	\label{fig:plate_diagram}
\end{figure}

\subsection{Linking CT to clinical outcomes}

Our model already leverages all 7 sets of measurements to estimate the idealized eCT. By adding additional hierarchy, we can learn about clinical outcomes while accounting for uncertainty in eCT itself. We achieve this with a linear mixed effects model. To predict MMSE score, we control for potential confounders by including age, gender, and diagnostic cohort as covariates. Furthermore, we think of all possible subjects with the same set of covariate values as a group. The fixed effects predict the mean at each time point for this group. But each subject also has “random effects,” or subject-specific effects: the subject's outcomes differ from the group mean by some amount, either as a function of time (random slope) or not (random intercept). Random effects allow for individual characteristics not captured by the covariates in the model. They apply to every observation for a given subject. They share a common distribution whose parameters are themselves to be estimated, and this constrains the random effects to be more similar than they would be if estimated separately. The remaining differences between predicted and observed outcomes arise as independent identically distributed random errors.

Let $MMSE_{ij}$ be the MMSE score of subject $i$ at visit $j$. Each subject is in one diagnostic group: CN, MCI, or AD. CN is the reference category, and $MCI_i$ and $AD_i$ are indicators. $AGE_i$ is the subject’s age at the initial visit; it is not a function of time. Female is the reference category, and $MALE_i$ is an indicator. $eCT_{ij}$ is not an observation but a parameter, our inference of idealized cortical thickness from the 7 pipeline measurements (Section \ref{sec:pipeline_descriptions}). $YEARS_{ij}$ is the time, in years, between the initial visit and visit $j$. The parameters $\alpha_{0i}$ and $\alpha_{1i}$ denote the random intercept and slope, and $\epsilon_{ij}$ the error.

To be clear, we have 2449 observations each in $MMSE$ and $YEARS$, and 663 each in $MCI$, $AD$, and $AGE$. We use them, along with the pipeline-specific eCT measurements mediated through the $eCT$ variable, to estimate 663 parameters each in $\alpha_{0i}$ and $\alpha_{1i}$, the hyperparameters $\lambda_0$ and $\lambda_1$, and the 9 parameters in $\beta$. Our priors are weakly informative, giving more weight to plausible values than implausible ones, but allowing most of the weight to come from the data:
\begin{gather*}
MMSE_{ij}=\beta_0 + \beta_{mci}MCI_i + \beta_{ad}AD_i + \beta_{age}AGE_i + \beta_{male}MALE_i + \beta_{ct}CT_{ij} + \alpha_{0i} \\ + (\beta_1 + \beta_{mci \times t}MCI_i + \beta_{ad \times t}AD_i + \alpha_{1i}) * YEARS_{ij} + \epsilon_{ij}, \\
\alpha_{0i} \sim N(0, \lambda_0), \quad
\alpha_{1i} \sim N(0, \lambda_1), \quad
\epsilon_{ij} \sim N(0, \sigma). \\
\beta_0 \sim N(15,15), \quad
\lambda_0 \sim N(0,10), \quad
\beta_1 \sim N(0,5), \quad
\lambda_1 \sim N(0,10) \\
\beta_{mci} \sim N(0,10), \quad
\beta_{ad} \sim N(0, 10), \quad
\beta_{age} \sim N(0,10), \quad
\beta_{ct} \sim N(0,10) \\
\beta_{mci \times t} \sim N(0,10), \quad
\beta_{ad \times t} \sim N(0, 10), \quad
\sigma \sim N^{+}(0,1).
\end{gather*}

Typically, in models like this one, we estimate the correlation between the random slopes and random intercepts. In our data, for example, it is conceivable that subjects with lower initial eCT than average are more likely to have rapidly progressing dementia, and to be losing eCT faster than average. This would show up as a positive correlation. Or it might be that those with greater initial eCT than average have more to lose, and are losing it faster. This would show up as a negative correlation. From an alternative version of the model where
\begin{gather*}
	\begin{bmatrix} \alpha_{0i} \\ \alpha_{1i} \end{bmatrix} \sim
	N_2 \left(
	\begin{bmatrix} 0 \\ 0 \end{bmatrix},
	\begin{bmatrix} \lambda_{11} & \lambda_{12} \\ \lambda_{12} &  \lambda_{22} \end{bmatrix}
	\right),
\end{gather*}
we found that $\lambda_{12} \approx 0$, allowing us to simplify our model by treating the random effects as independent. Figure \ref{fig:plate_diagram} visualizes the complete model as a plate diagram.

\section{Computation}

The model has 3829 explicit parameters: idealized eCT for each of 2449 observations; a random slope $\alpha_{0i}$ and intercept $\alpha_{1i}$ for 663 subjects; the means $\alpha_{0}, \alpha_{1}$ and standard deviations $\lambda_0, \lambda_1$ of the random slopes and intercepts; the 7 regression coefficients $\beta$ and within subject error scale $\sigma$; the offsets $\phi$, scale parameters $\tau$, and degrees of freedom $\nu$ for each of 7 pipelines; the 21 correlation coefficients $\rho_{kk'}$ between pipeline errors. Implementing the mixture form of the NECT distribution requires estimating $\chi^{2}_{\nu_k n}$, with $k=1:7$ and $n=1:2449$, adding 17143 latent variables.

Fitting this model yields the joint probability, given the data and the assumptions of the model, of every possible combination of parameter values. Remarkably, this means that we can infer all the parameters simultaneously—in this case many thousands of them—and every datum is used to estimate every parameter. Not only the pipeline measurements, but also age, sex, diagnosis, MMSE, and even the visit times inform our estimate of $eCT$.

We use the R programming language \citep{rbase, rstudio}, supplemented by the package dplyr \citep{tidyverse} for all data management, and by the package pracma \citep{pracma} for advanced mathematical operations. To specify and fit the model, we use Stan, an open-source Bayesian software application, interfacing with R through the rstan package \citep{rstan}. We include our model script as an appendix, and the Stan file and all R scripts at \url{https://github.com/andrewjholbrook/pipeline_integration_manu}. Stan uses Hamiltonian Monte Carlo, a flavor of Markov Chain Monte Carlo \citep{neal2011mcmc, hoffman2014no} to draw samples—each of which is a complete vector of parameters—from the joint posterior distribution. An advanced extension of HMC called the No U-Turn Sampler (NUTS) improves sampling efficiency \citep{hoffman2014no}. As the number of MCMC samples increases, the distribution of MCMC samples converges to the actual joint posterior distribution. Fortunately, the marginal distribution of each parameter in the MCMC samples converges to its actual marginal posterior distribution as well. We ran 100 chains in parallel with a minimum chain length of 36,000 iterations, for a total of 11,823,200 iterations, which we thinned by a factor of 200 to get 59,116 iterations. Effective sample sizes are all greater than 500. See Appendix \ref{sec:appendixA} for the Stan code for the model.

\begin{table}[htp!]
	\centering
	\begin{tabular}{lllrrr}
		\toprule
		parameter & symbol & element & mean & CI low & CI high \\
		\midrule
		pipeline error offset & $\phi$   & FSCross 		 & -1.02 & -1.10 & -0.94 \\
		&& FSLong 		 & -1.00 & -1.08 & -0.92 \\
		&& ANTsCross 	 & 0.92  & 0.83  & 1.01 \\
		&& ANTsNative 	 & 0.21  & 0.12  & 0.30 \\
		&& ANTsSST 		 & 0.23  & 0.14  & 0.32 \\
		&& ANTsXNetCross & -0.51 & -0.59 & -0.42 \\
		&& ANTsXNetLong  & 0.24  & 0.15  & 0.32 \\
		pipeline error scale & $\tau$    & FSCross 	     & 0.21  & 0.16  & 0.28 \\
		&& FSLong 		 & 0.24  & 0.17  & 0.31 \\
		&& ANTsCross 	 & 1.06  & 0.99  & 1.13 \\
		&& ANTsNative 	 & 1.23  & 1.17  & 1.30 \\
		&& ANTsSST 		 & 1.27  & 1.21  & 1.34 \\
		&& ANTsXNetCross & 0.97  & 0.91  & 1.03 \\
		&& ANTsXNetLong  & 0.79  & 0.73  & 0.85 \\
		pipeline error df & $\nu$ 		 & FSCross 		 & 17.43 & 5.20  & 77.25 \\
		&& FSLong 		 & 6.06  & 4.03  & 9.18 \\
		&& ANTsCross 	 & 15.66 & 10.65 & 23.88 \\
		&& ANTsNative 	 & 105.02& 57.94 & 185.69 \\
		&& ANTsSST 		 & 54.78 & 33.74 & 91.47 \\
		&& ANTsXNetCross & 35.95 & 21.77 & 59.62 \\
		&& ANTsXNetLong  & 13.01 & 9.63  & 17.60 \\
		random effect mean & $\alpha$    & intercept 	 & 22.29 & 19.71 & 24.88 \\
		&& slope 	 	 & 0.01  & -0.25 & 0.26 \\
		random effect sd & $\lambda$     & intercept 	 & 1.61  & 1.48  & 1.75 \\
		&& slope 	 	 & 1.53  & 1.39  & 1.68 \\
		regression coefficient & $\beta$ & MCI  		 & -1.72 & -2.09 & -1.33 \\
		&& AD   		 & -4.86 & -5.35 & -4.36 \\
		&& age  		 & 0.01  & -0.01 &  0.04 \\
		&& male 	 	 & 0.00  & -0.30 &  0.31 \\
		&& eCT   	 	 & 0.75  & 0.57  &  0.94 \\
		&& MCI$\times$t  & -0.81 & -1.14 & -0.47 \\
		&& AD$\times$t   & -2.35 & -2.79 & -1.90 \\
		within-subject error & $\sigma$  &               & 1.48  & 1.42  & 1.54 \\
		\bottomrule
	\end{tabular}
	\caption{Posterior means and 95\% credible intervals (CI) for all parameters of interest. In the Bayesian framework, the posterior mean is an estimator with many desirable properties, e.g. minimizing the mean squared error, and one that stochastic methods can accurately estimate. Regression coefficients are significant if the credible intervals do not include 0. Pipelines are variants of FreeSurfer (FS) and Advanced Normalization Tools (ANTs). Cognitively normal (CN) is the baseline diagnostic category, and we have indicators for mildly cognitively impaired (MCI) and Alzheimer's disease (AD).}
	\label{table:mytable}
\end{table}

\section{Results}

\begin{figure}[htp!]
	\centering
	\includegraphics[width=0.7\linewidth]{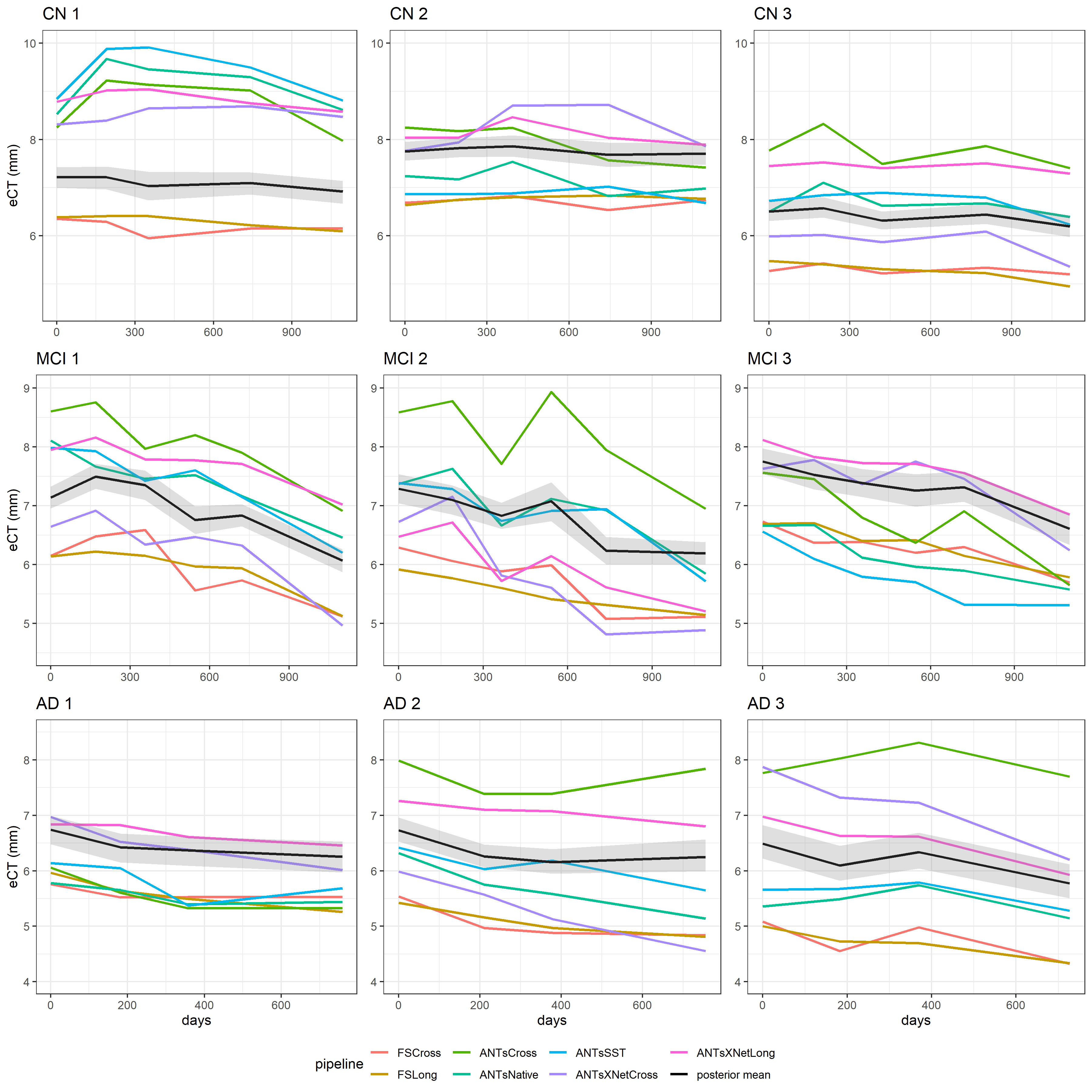}

	\caption{Profiles with posterior mean and 95\% credible interval of cortical thickness (eCT) for nine randomly selected individuals, three from each diagnostic category: cognitively normal (CN), mildly cognitively impaired (MCI), and Alzheimer's disease (AD). Pipelines are variants of FreeSurfer (FS) and Advanced Normalization Tools (ANTs). The interval is much narrower than the range of the pipelines, indicating an increase in precision. The posterior mean tends to run roughly parallel to the pipelines, but it dampens the fluctuations. It tends to fall near the midrange, but it also prefers initial values near 7.}
	\label{fig:model_10f_profiles}
\end{figure}

\begin{figure}[htp!]
	\centering
	\includegraphics[width=0.7\linewidth]{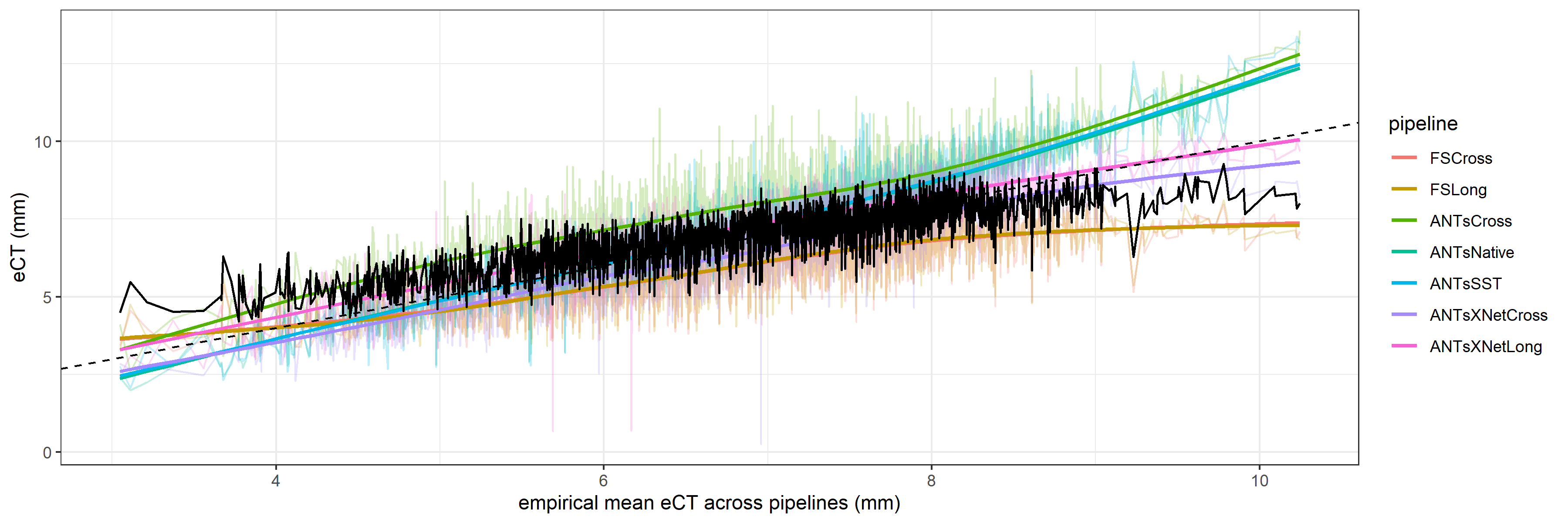}

	\caption{Posterior mean cortical thickness (eCT) as a function of empirical mean cortical thickness. Pipelines are variants of FreeSurfer (FS) and Advanced Normalization Tools (ANTs). The jaggedness is an artifact of arranging the observations by empirical mean. For the bulk of the data, it falls near the midrange of the pipeline estimates and tends to avoid extremes.}
	\label{fig:model_10f_spaghetti}
\end{figure}

\begin{figure}[htp!]
	\centering
	\includegraphics[width=0.7\linewidth]{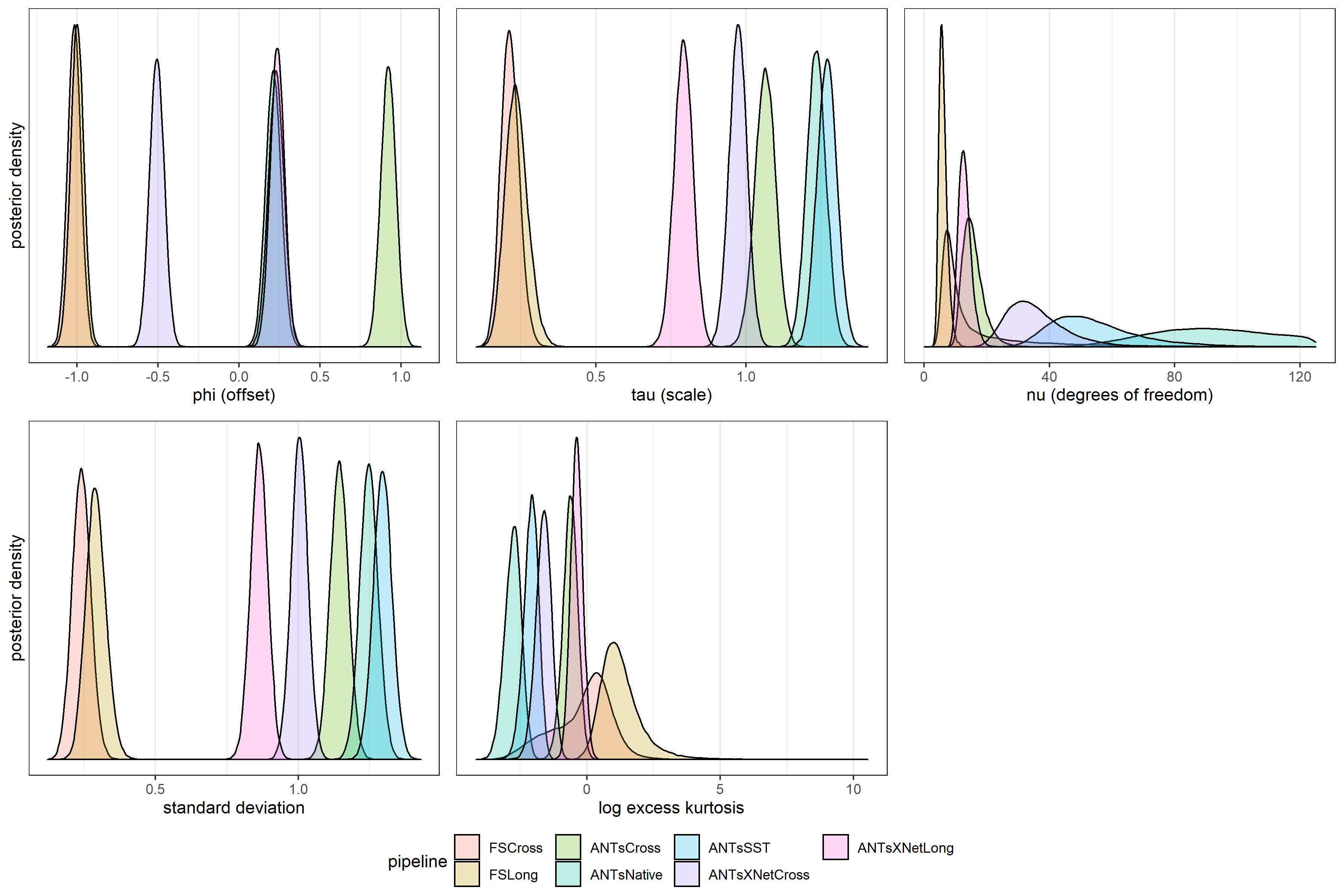}

	\caption{Posterior distributions for measurement error model and derived parameters describing cortical thickness. Among other things, they suggest that the FreeSurfer (FS) pipelines have greater offset than do Advanced Normalization Tools (ANTs) pipelines. FS pipelines also have smallest standard deviation and greatest kurtosis, a tradeoff discussed in Section \ref{sec:tdist}.}
	\label{fig:model_10f_posteriors}
\end{figure}

\begin{figure}[ht!]
	\centering
	\includegraphics[width=0.7\linewidth]{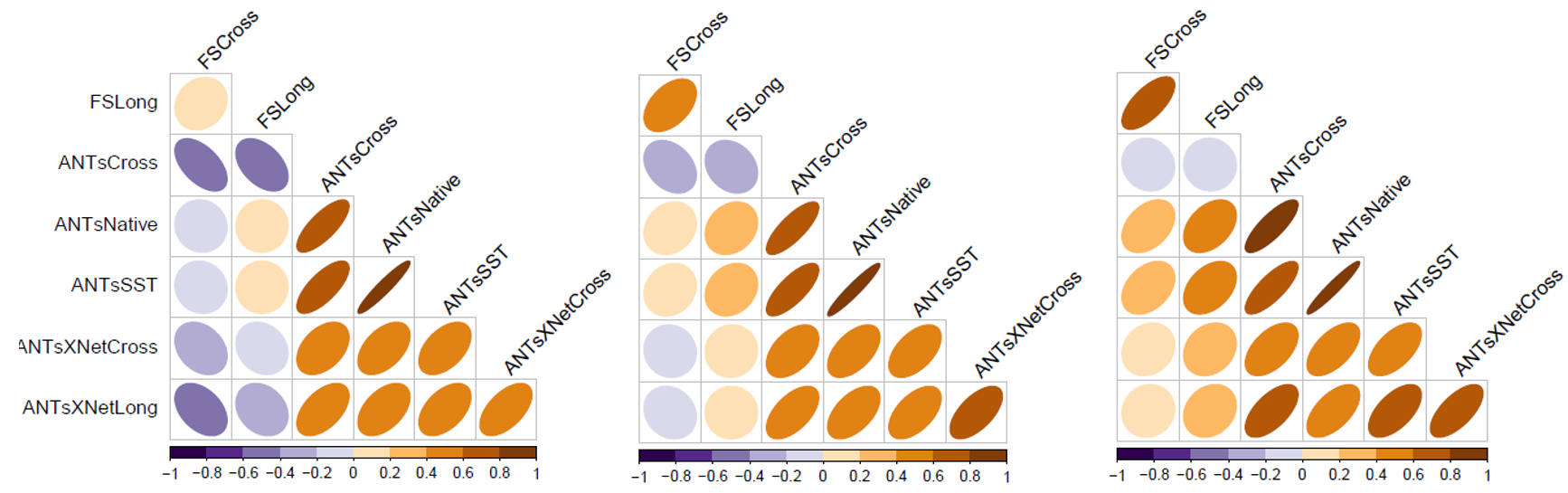}

	\caption{Pipeline error correlations. Left: posterior credible interval lower bound, center: posterior mean, right: posterior credible interval upper bound. Pipelines are variants of FreeSurfer (FS) and Advanced Normalization Tools (ANTs). We see strong positive correlations between the errors of FSCross and FSLong, and between ANTsNative, ANTsCross, and ANTsSST. We see significant negative correlations between the errors of the FreeSurfer pipelines and ANTsCross.}
	\label{fig:model_10f_error_cor_all}
\end{figure}

\begin{figure}[ht!]
	\centering
	\includegraphics[width=0.7\linewidth]{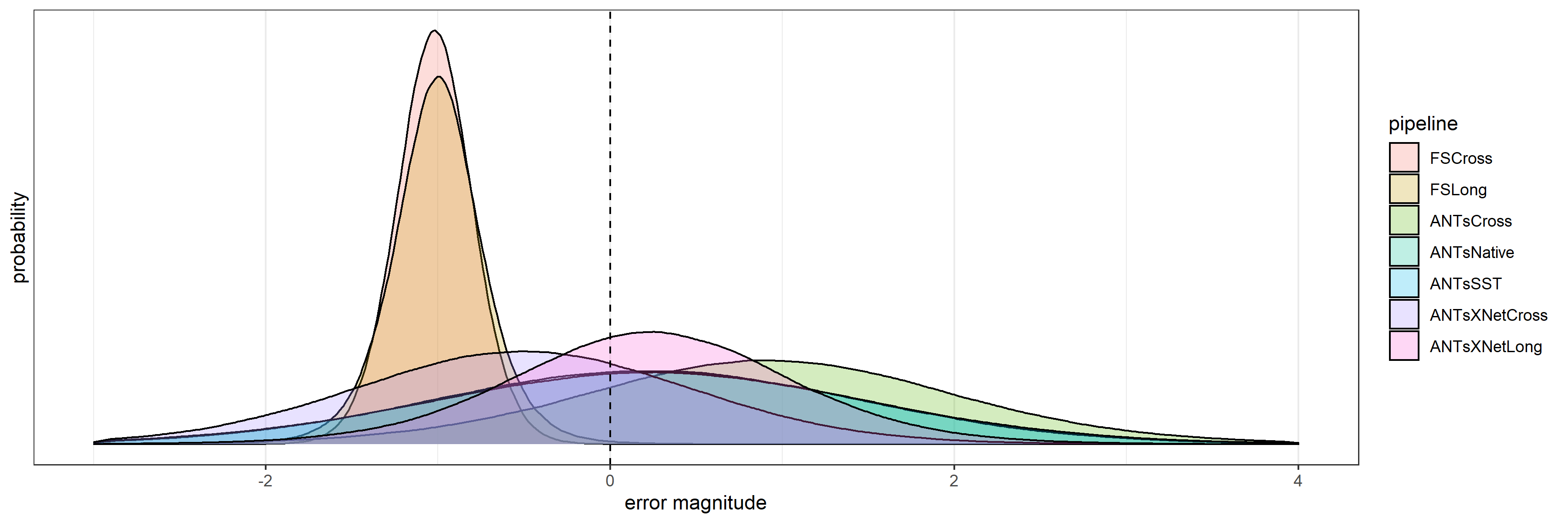}

	\caption{Using $t$ distributions with the posterior means of $\phi$ (offset), $\tau$ (scale), and $\nu$ (degrees of freedom), we plot the probability of different magnitude of cortical thickness measurement errors, in mm, for the different pipelines. This plot does not reveal error correlations. Compared to the Advanced Normalization Tools (ANTs) pipelines, the FreeSurfer pipelines show smaller standard deviations but also greater offsets.}
	\label{fig:error_dist_plot}
\end{figure}

\begin{figure}[ht!]
	\centering
	\includegraphics[width=0.7\linewidth]{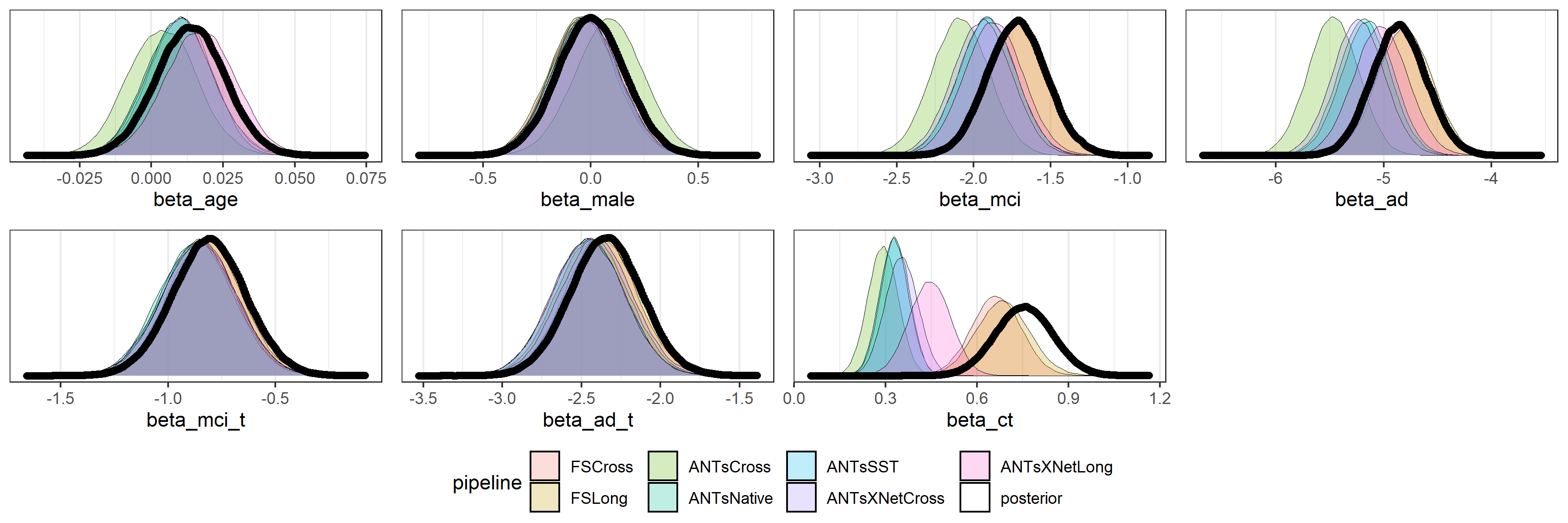}

	\caption{Posterior distributions for regression coefficients in longitudinal model for the Mini Mental State Examination (MMSE), estimated from individual pipeline data and from combined data. Pipelines are variants of FreeSurfer (FS) and Advanced Normalization Tools (ANTs). Mild cognitive impairment (MCI) has a strong and significant negative effect on both intercept and slope of MMSE score over time. This is even more true for Alzheimer's disease (AD). Cortical thickness (eCT) has a significant positive effect, clearer in the combined estimate than in any of the individual pipeline estimates.}
	\label{fig:single_pipeline_combined_beta_posteriors}
\end{figure}

Table \ref{table:mytable} gives the posterior mean and 95\% credible intervals for all parameters of interest. We now explore their implications by examining several visualizations. In Figure \ref{fig:model_10f_profiles}, we add to the profile plots a profile for the posterior mean eCT, with a 95\% credible interval. While our posterior estimate tends to run parallel to the individual pipeline measurements, it also tends to follow a smoother trajectory, dampening large fluctuations. It usually falls near the midrange of the measurements, except when the measurements are unusually large or small. In Figure \ref{fig:model_10f_spaghetti}, we add to the line plot a line for the posterior mean. Again we have posterior eCT typically near the midrange of the pipeline measurements for the bulk of the data, but it tends to avoid extremes. Individual differences, which we model as random effects, show up in profile plots. The posterior mean for $\alpha_0$, the mean of the random intercepts, is 22.29 with credible interval $(19.71, 24.88)$. The posterior mean for $\lambda_0$, the standard deviation of the random intercepts, is  1.61 with credible interval $(1.48, 1.75)$. The posterior mean for $\alpha_1$ is 0.01 with credible interval $(-0.25, 0.26)$. The posterior mean for $\lambda_1$, the standard deviation of the random slopes, is 1.53 with credible interval $(1.39, 1.68)$.

The posterior estimates for the offsets $\phi$, listed in the Table \ref{table:mytable} and visualized in Figure \ref{fig:model_10f_posteriors}, reflect these patterns. They are as follows: FS\-Cross (-1.02; -1.10, -0.94), FS\-Long (-1.00; -1.08, -0.92), ANTs\-Cross (0.92; 0.83, 1.01), ANTs\-Native (0.21; 0.12, 0.30), ANTs\-SST (0.23; 0.14, 0.32), ANTsX\-NetCross (-.51; -0.59, -0.42), ANTsX\-NetLong (0.24; 0.16, 0.28). The crossing effect in Figure \ref{fig:prior_spaghetti} may be part of why Native and SST have the smallest estimated offsets, and why NetLong is given a similar offset to Native and SST, despite not being parallel. Offsets are estimated fairly precisely: posteriors are narrow. As we would expect from our visualization of the data, in offset FSCross $\approx$ FSLong, and ANTsNative $\approx$ ANTsSST.

The scale parameters $\tau$ are as follows: FSCross (0.21; 0.16, 0.28), FSLong (0.24; 0.17, 0.31), ANTsCross (1.06; 0.99, 1.13), ANTsNative (1.23; 1.17, 1.30), ANTsSST (1.27; 1.21, 1.34), ANTsX\-NetCross (0.97; 0.91, 1.03), ANTsX\-NetLong (0.79; 0.73, 0.85). The scale parameter is different from the standard deviation; it is the horizontal scale factor of the error distribution compared to the standard $t$ distribution. But both parameters tell a similar story. Smaller values of each indicate more precision, a smaller spread of estimates. FSCross and FSLong are most precise by this metric.

The degrees of freedom parameters $\nu$ are as follows: FSCross (17.43; 5.20, 77.25), FSLong (6.06; 4.03, 9.18), ANTsCross (15.66; 10.65, 23.88), ANTsNative (105.02; 57.94, 185.69), ANTsSST (54.78; 33.74, 91.47), ANTsX\-NetCross (35.95; 21.77, 59.62), ANTsX\-NetLong (13.01; 9.63, 17.60). Since raw $\nu$ is rather hard to interpret, we also include log excess kurtosis in Figure \ref{fig:model_10f_posteriors}, a derived parameter. Larger values indicate heavier tails, which indicate a higher frequency of outliers. It is not a coincidence that FSCross and FSLong, which have the lowest spread, also have the highest kurtosis. With a $t$ distribution, the model partitions deviation into both.

Figure \ref{fig:model_10f_error_cor_all} shows posterior estimates (left: CI lower bound, center: mean, right: CI upper bound) for each pairwise correlation between pipeline errors. We find strong positive correlations between the FS pipelines $(0.47; 0.17, 0.68)$, and especially between ANTsCross and ANTsNative $(0.79; 0.77, 0.81)$, ANTsCross and ANTsSST 3,5 $(0.76; 0.73, 0.78)$, and ANTsNative and ANTsSST 4,5 $(0.95; 0.94, 0.96)$. We find negative correlations between FSCross and ANTsCross $(-0.32; -0.54, -0.09)$ and between FSLong and ANTsCross 2,3 $(-0.24; -0.45, -0.03)$.

Figure \ref{fig:error_dist_plot} visualizes the marginal probability distribution of errors for each pipeline, according to the parameters estimated by our model. The FreeSurfer pipelines show smaller standard deviations but greater offsets. However, Appendix \ref{sec:appendixB} shows that these marginal representations fail to capture important correlation structure and the non-elliptical nature of the posterior multivariate measurement error distribution.

In Figure \ref{fig:single_pipeline_combined_beta_posteriors}, the heavy lines are the posterior estimates of the regression coefficients, which quantify the association of the covariates with MMSE score. Our results provide no evidence that, conditioning on all the other variables, age $(0.01; -0.01, 0.04)$ or sex $(0.00; -0.30, 0.31)$ makes any difference. However, compared to cognitively normal subjects, and on average, those with mild cognitive impairment have lower MMSE score (-1.72; -2.09, -1.33), and those with Alzheimer's disease lower still $(-4.86; -5.35, -4.36)$. The interaction coefficients indicate that those with MCI decline faster $(-0.81; -1.14, -0.47)$, and those with AD faster still $(-2.35; -2.79, -1.90)$. Finally, each millimeter of eCT is associated with a gain of a half point to a whole point on the assessment $(0.75; 0.57, 0.94)$.

We include coefficient posterior distributions based on each individual pipeline to demonstrate how combining the pipelines affects our estimates. The difference is most obvious with  $\beta_{eCT}$. The posterior distribution is wider because, unlike the individual models, the combined model captures the genuine uncertainty in the data and reflects it with uncertainty in the conclusions. Yet, interestingly, the effect size is also greater, suggesting that a real association, muffled by measurement error, has been recovered as we improved the accuracy of the data. \citet{carroll2006measurement} discusses how, in the simple linear regression setting, measurement error for covariates leads to attenuation, or the bias toward zero of coefficient estimates. Something similar may be occurring here.

\section{Discussion}

We set ourselves an ambitious goal in this project. Given conflicting sets of measurements, in the absence of ground truth and without the possibility of external validation, attempted to make the most accurate possible estimates of EC thickness \emph{and} to gauge how accurate these estimates are. We have reasons to believe that we have made progress.

First, our model incorporates a large amount of data. The eCT measurements, several per individual, come from 7 different pipelines from 2 different paradigms (FS and ANTs), comprising a variety of image processing algorithms and computational methods. Since we estimate all parameters together to find the joint posterior probability of each combination of values, every data point contributes to the accuracy of every parameter estimate.

Second, we make effective use of all this data with a complex and carefully specified model. We allow pipeline errors to have different offsets (systematic error) and different spreads (random error). The random components are not necessarily Gaussian; they may have different scales and different tail thicknesses to account for frequency of outliers. We account for error correlation, so that duplicated measurements are not given undue weight. As we have multiple observations per subject, we structure them longitudinally, taking time trends into account, rather than just pooling them. Furthermore, each subject has a ``random" (subject-specific) slope and intercept to allow for systematic individual differences not accounted for by the covariates and to reduce within-subject error correlation. Each subject's data contributes to estimation of hyperparameters. This hierarchical structure allows sharing of information across subjects.

Third, with our fully Bayesian approach, we have a principled and accurate way to propagate the uncertainty due to measurement error, as well as uncertainty from other sources, into our model for the relationship of eCT and cognitive capacity over time. Because fitting such a complex model is difficult, we use state of the art tools for computation. The Stan environment for Hamiltonian Monte Carlo (HMC), with the No U-Turn Sampler algorithm, generates samples of complete parameter vectors from the joint posterior. These samples converge to the true joint posterior. It is straightforward to find the marginal distribution of a single parameter (or subset): just take the corresponding element (or elements) of each sample vector. We report posterior means, but any function of the posterior can be an estimator. Credible intervals are just quantiles of the posterior.

Conceivably, an approach like ours might one day be a stage in a larger pipeline. ADNI and other initiatives create images. FreeSurfer and ANTs do extraction, registration, labeling, and other processing, and among other things deliver cortical thickness estimates. A Bayesian hierarchical model like this one, tuned specifically to each data set, would synthesize the estimates and quantify the uncertainty.

However, we recognize that many questions remain. To begin with, we synthesize seven different eCT pipelines. But while this set is a good representation, it is not the only choice. In the future, we may experiment with different sets. Including more might give us more information, or perhaps including fewer might eliminate redundancy and simplify computation. We may also explore different model specifications. Our prior on eCT is intended to be regularizing, reining in large errors, but it may have too strong an effect on the highest and lowest actual eCT values. We could let it depend on certain covariates, like diagnostic category. For the eCT measurement error, the non-elliptically contoured $t$ distribution gives us a lot of power and flexibility, but with Stan's ability to code scale mixtures, we have a wide variety of choices. Here we model the offsets (systematic pipeline-specific errors) $\phi$ as constants, but there is an argument for allowing them to depend linearly on the magnitude of eCT, e.g.,
 \begin{center}
 	$\widehat{eCT} = eCT + (\phi_0 + \phi_1 eCT)$ + $\epsilon$.\\
 \end{center}
We have MMSE as our outcome, and we model the scores as fixed--without measurement error--even though psychometric tools are probably quite a bit less precise than neuroimaging tools. The ADNI-1 data set includes CDR SOB scores as well, so we could model a bivariate outcome. And we could use a more sophisticated within-subject error model for the longitudinal component than random slope and intercept.

Ultimately, we hope to extend the model to jointly incorporate the measurements of many, if not all, of the 62 brain regions of the DKT atlas. We could look at new layers of measurement error correlation--the way the same pipeline performs on different structures. And we could get a picture of how the Alzheimer's brain as a whole develops over time.

\appendix

\section{Stan code}\label{sec:appendixA}

\begin{verbatim}

data {

	int<lower=1>       I; // number of individuals
	int<lower=1>       N; // number of observations, total
	int<lower=1>       K; // number of pipelines
	int<lower=1>       ID[N]; // 1-d array of each individual id
	vector[N]          YEARS; // days since initial visit
	vector[N]          INITIAL_AGE; // age at first visit
	vector[N]          MALE; // male indicator
	vector[N]          MCI; // MCI indicator
	vector[N]          AD; // AD indicator
	vector[N]          MM_SCORE; // score on cognitive test
	matrix[N, K]       Y; // observation of L + R ERC thickness

}

transformed data {
	matrix[N, 6] X;

	X[,1] = MCI;
	X[,2] = AD;
	X[,3] = INITIAL_AGE;
	X[,4] = MALE;
	X[,5] = MCI .* YEARS;
	X[,6] = AD .* YEARS;
}

parameters {

	vector[I]               Z0;
	vector[I]               Z1;
	vector[I]               Z2;

	real                    alpha_0;
	real                    alpha_1;
	real<lower=0>           lambda_0;
	real<lower=0>           lambda_1;
	real                    beta_mci;
	real                    beta_ad;
	real                    beta_age;
	real                    beta_male;
	real                    beta_ct;
	real                    beta_mci_t;
	real                    beta_ad_t;
	real<lower=0>           sigma;
	vector<lower=0>[N]      CT;
	vector[K]               offset;
	vector<lower=0>[K]      tau; // scale
	vector<lower=0>[K]      nu; // df
	cholesky_factor_corr[K] L_Omega; // Omega is CT error correlation matrix
	vector<lower=0>[K]      chisqnu[N];

}

transformed parameters {

}

model{

	vector[6] beta = [
	beta_mci,
	beta_ad,
	beta_age,
	beta_male,
	beta_mci_t,
	beta_ad_t
	]';

	vector[I] alpha_0_subject = alpha_0 + lambda_0 * Z0;
	vector[I] alpha_1_subject = alpha_1 + lambda_1 * Z1;

	Z0 ~ normal(0,1);
	Z1 ~ normal(0,1);
	Z2 ~ normal(0,1);

	alpha_0 ~ normal(15, 15);
	lambda_0 ~ normal(0, 10);
	alpha_1 ~ normal(0, 5);
	lambda_1 ~ normal(0, 10);

	beta_mci ~ normal(0, 10);
	beta_ad ~ normal(0, 10);
	beta_age ~ normal(0, 10);
	beta_male ~ normal(0, 10);
	beta_ct ~ normal(0, 10);
	beta_mci_t ~ normal(0, 10);
	beta_ad_t ~ normal(0, 10);
	sigma ~ normal(0, 1);

	CT ~ normal(7, 2);
	offset ~ normal(0, 3);
	tau ~ normal(0, 1);
	nu ~ exponential(1.0/30.0);
	L_Omega ~ lkj_corr_cholesky(2);

	{
		vector[K] q[N];
		vector[K] sqrtinvq_tau[N];

		for(n in 1:N){
			chisqnu[n] ~ chi_square(nu);
			q[n] = chisqnu[n] ./ nu;
			sqrtinvq_tau[n] = tau ./ sqrt(q[n]);
			Y[n] ~ multi_normal_cholesky(CT[n] + offset, diag_pre_multiply(sqrtinvq_tau[n], L_Omega));
		}
	}

	MM_SCORE ~ normal_id_glm(
	X,
	beta_ct * CT + alpha_0_subject[ID] + alpha_1_subject[ID] .* YEARS,
	beta,
	sigma
	);

}

generated quantities {

	matrix[K, K] Omega = L_Omega * L_Omega';

}

\end{verbatim}

\section{Inferring a non-elliptical error model}\label{sec:appendixB}

\begin{figure}[ht!]
	\centering
	\includegraphics[width=0.7\linewidth]{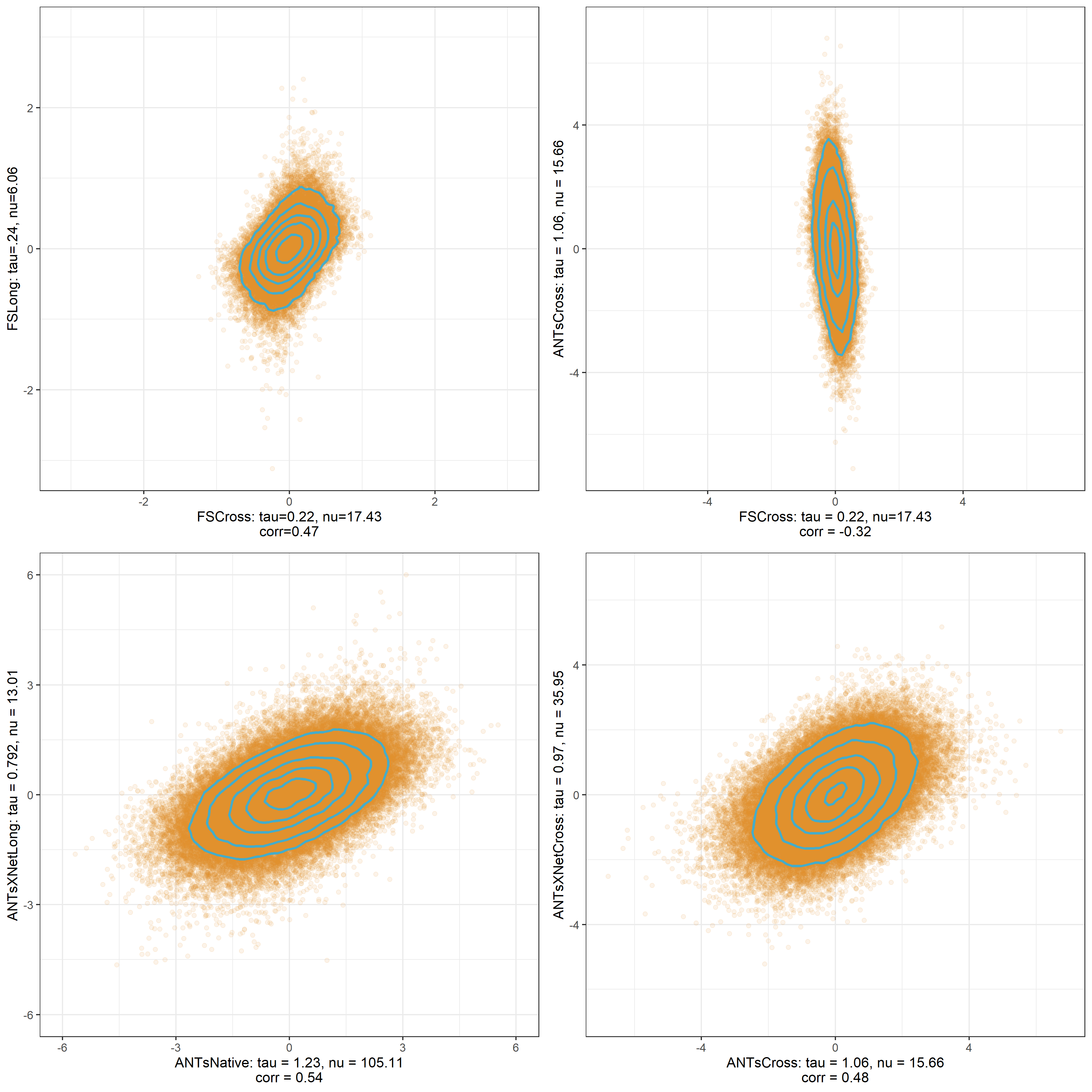}
	\caption{We visualize the non-elliptically contoured $t$ (NECT) distribution by generating samples from bivariate distributions, using two pipelines at a time. Pipelines are variants of FreeSurfer (FS) and Advanced Normalization Tools (ANTs). As parameters, we use different pipeline posterior means.}
	\label{fig:nect_plots}
\end{figure}

Figure \ref{fig:nect_plots} gives examples of the shape of the NECT distribution in the bivariate case, according to some parameters estimated by our model. In the upper left plot, we have similar $\tau$ (scale), different $\nu$ (degrees of freedom), and positive correlation. Since $\tau$'s are similar, the distributions have similar spread for the most part--the pattern is nearly circular, only slightly elliptical--but the distribution with smaller $\nu$ has much further outliers. Positive correlation accounts for the tilt. In the upper right plot, we have similar $\nu$, different $\tau$, and negative correlation. The difference in $\tau$ overwhelms the difference is $\nu$ here and stretches the entire distribution, including the tails. In the bottom left plot, we have different $\tau$ and $\nu$ and positive correlation. With fairly large $\nu$ in both variables, $\tau$ accounts for most of the difference, and we don't see far outliers. In the bottom right plot, we have different $\tau$ and $\nu$ and positive correlation. It creates a similar pattern to the bottom left plot.

\bibliographystyle{sysbio}
\bibliography{refs}

\begin{thebibliography}{23}
\providecommand{\natexlab}[1]{#1}
\providecommand{\selectlanguage}[1]{\relax}
\providecommand{\bibAnnoteFile}[1]{%
  \IfFileExists{#1}{\begin{quotation}\noindent\textsc{Key:} #1\\
  \textsc{Annotation:}\ \input{#1}\end{quotation}}{}}
\providecommand{\bibAnnote}[2]{%
  \begin{quotation}\noindent\textsc{Key:} #1\\
  \textsc{Annotation:}\ #2\end{quotation}}

\bibitem[{Barnard et~al.(2000)Barnard, McCulloch, and
  Meng}]{barnard2000modeling}
Barnard, J., R.~McCulloch, and X.-L. Meng. 2000. Modeling covariance matrices
  in terms of standard deviations and correlations, with application to
  shrinkage. Statistica Sinica Pages~1281--1311.
\bibAnnoteFile{barnard2000modeling}

\bibitem[{Borchers(2021)}]{pracma}
Borchers, H.~W. 2021. pracma: practical numerical math functions. R package
  version 2.3.3.
\bibAnnoteFile{pracma}

\bibitem[{Carroll et~al.(2006)Carroll, Ruppert, Stefanski, and
  Crainiceanu}]{carroll2006measurement}
Carroll, R.~J., D.~Ruppert, L.~A. Stefanski, and C.~M. Crainiceanu. 2006.
  Measurement error in nonlinear models: a modern perspective. Chapman and
  Hall/CRC.
\bibAnnoteFile{carroll2006measurement}

\bibitem[{Cleveland and Devlin(1988)}]{cleveland1988locally}
Cleveland, W.~S. and S.~J. Devlin. 1988. Locally weighted regression: an
  approach to regression analysis by local fitting. Journal of the American
  statistical association 83:596--610.
\bibAnnoteFile{cleveland1988locally}

\bibitem[{Das et~al.(2009)Das, Avants, Grossman, and Gee}]{das2009registration}
Das, S.~R., B.~B. Avants, M.~Grossman, and J.~C. Gee. 2009. Registration based
  cortical thickness measurement. Neuroimage 45:867--879.
\bibAnnoteFile{das2009registration}

\bibitem[{et~al.(2019)}]{tidyverse}
et~al., H.~W. 2019. Welcome to the tidyverse. Journal of Open Source Software
  4:1686.
\bibAnnoteFile{tidyverse}

\bibitem[{Fischl(2012)}]{fischl2012freesurfer}
Fischl, B. 2012. Freesurfer. Neuroimage 62:774--781.
\bibAnnoteFile{fischl2012freesurfer}

\bibitem[{Gelman et~al.(1995)Gelman, Carlin, Stern, and
  Rubin}]{gelman1995bayesian}
Gelman, A., J.~B. Carlin, H.~S. Stern, and D.~B. Rubin. 1995. Bayesian data
  analysis. Chapman and Hall/CRC.
\bibAnnoteFile{gelman1995bayesian}

\bibitem[{Hoffman et~al.(2014)Hoffman, Gelman et~al.}]{hoffman2014no}
Hoffman, M.~D., A.~Gelman, et~al. 2014. The no-u-turn sampler: adaptively
  setting path lengths in hamiltonian monte carlo. J. Mach. Learn. Res.
  15:1593--1623.
\bibAnnoteFile{hoffman2014no}

\bibitem[{Holbrook et~al.(2020)Holbrook, Tustison, Marquez, Roberts, Yassa,
  Gillen, and {\S}}]{holbrook2020anterolateral}
Holbrook, A.~J., N.~J. Tustison, F.~Marquez, J.~Roberts, M.~A. Yassa, D.~L.
  Gillen, and A.~D. N.~I. {\S}. 2020. Anterolateral entorhinal cortex thickness
  as a new biomarker for early detection of alzheimer's disease. Alzheimer's \&
  Dementia: Diagnosis, Assessment \& Disease Monitoring 12:e12068.
\bibAnnoteFile{holbrook2020anterolateral}

\bibitem[{Jack~Jr et~al.(2008)Jack~Jr, Bernstein, Fox, Thompson, Alexander,
  Harvey, Borowski, Britson, L.~Whitwell, Ward et~al.}]{jack2008alzheimer}
Jack~Jr, C.~R., M.~A. Bernstein, N.~C. Fox, P.~Thompson, G.~Alexander,
  D.~Harvey, B.~Borowski, P.~J. Britson, J.~L.~Whitwell, C.~Ward, et~al. 2008.
  The alzheimer's disease neuroimaging initiative (adni): Mri methods. Journal
  of Magnetic Resonance Imaging: An Official Journal of the International
  Society for Magnetic Resonance in Medicine 27:685--691.
\bibAnnoteFile{jack2008alzheimer}

\bibitem[{Jiang and Ding(2016)}]{jiang2016robust}
Jiang, Z. and P.~Ding. 2016. Robust modeling using non-elliptically contoured
  multivariate t distributions. Journal of Statistical Planning and Inference
  177:50--63.
\bibAnnoteFile{jiang2016robust}

\bibitem[{Kruschke(2014)}]{kruschke2014doing}
Kruschke, J. 2014. Doing bayesian data analysis: A tutorial with r, jags, and
  stan .
\bibAnnoteFile{kruschke2014doing}

\bibitem[{Mueller et~al.(2005)Mueller, Weiner, Thal, Petersen, Jack, Jagust,
  Trojanowski, Toga, and Beckett}]{mueller2005ways}
Mueller, S.~G., M.~W. Weiner, L.~J. Thal, R.~C. Petersen, C.~R. Jack,
  W.~Jagust, J.~Q. Trojanowski, A.~W. Toga, and L.~Beckett. 2005. Ways toward
  an early diagnosis in alzheimer’s disease: the alzheimer’s disease
  neuroimaging initiative (adni). Alzheimer's \& Dementia 1:55--66.
\bibAnnoteFile{mueller2005ways}

\bibitem[{Neal et~al.(2011)}]{neal2011mcmc}
Neal, R.~M. et~al. 2011. Mcmc using hamiltonian dynamics. Handbook of markov
  chain monte carlo 2:2.
\bibAnnoteFile{neal2011mcmc}

\bibitem[{Petersen et~al.(2010)Petersen, Aisen, Beckett, Donohue, Gamst,
  Harvey, Jack, Jagust, Shaw, Toga et~al.}]{petersen2010alzheimer}
Petersen, R.~C., P.~Aisen, L.~A. Beckett, M.~Donohue, A.~Gamst, D.~J. Harvey,
  C.~Jack, W.~Jagust, L.~Shaw, A.~Toga, et~al. 2010. Alzheimer's disease
  neuroimaging initiative (adni): clinical characterization. Neurology
  74:201--209.
\bibAnnoteFile{petersen2010alzheimer}

\bibitem[{{R Core Team}(2021)}]{rbase}
{R Core Team}. 2021. R: A language and environment for statistical computing.
\bibAnnoteFile{rbase}

\bibitem[{Rodgers(2002)}]{rodgers2002alzheimer}
Rodgers, A.~B. 2002. Alzheimer's disease: unraveling the mystery vol.~1.
  National Institutes of Health.
\bibAnnoteFile{rodgers2002alzheimer}

\bibitem[{{RStudio Team}(2020)}]{rstudio}
{RStudio Team}. 2020. Rstudio: integrated development environment for r.
\bibAnnoteFile{rstudio}

\bibitem[{Sahyouni et~al.(2017)Sahyouni, Brown, and
  Chen}]{sahyouni2017alzheimer}
Sahyouni, R., N.~Brown, and J.~Chen. 2017. Alzheimer’s disease decoded: The
  history, present, and future of Alzheimer’s disease and dementia. World
  Scientific.
\bibAnnoteFile{sahyouni2017alzheimer}

\bibitem[{{Stan Development Team}(2020)}]{rstan}
{Stan Development Team}. 2020. {RStan}: the {R} interface to {Stan}. R package
  version 2.21.2.
\bibAnnoteFile{rstan}

\bibitem[{Tombaugh and McIntyre(1992)}]{tombaugh1992mini}
Tombaugh, T.~N. and N.~J. McIntyre. 1992. The mini-mental state examination: a
  comprehensive review. Journal of the American Geriatrics Society 40:922--935.
\bibAnnoteFile{tombaugh1992mini}

\bibitem[{Tustison et~al.(2019)Tustison, Holbrook, Avants, Roberts, Cook,
  Reagh, Duda, Stone, Gillen, Yassa et~al.}]{tustison2019longitudinal}
Tustison, N.~J., A.~J. Holbrook, B.~B. Avants, J.~M. Roberts, P.~A. Cook, Z.~M.
  Reagh, J.~T. Duda, J.~R. Stone, D.~L. Gillen, M.~A. Yassa, et~al. 2019.
  Longitudinal mapping of cortical thickness measurements: An alzheimer’s
  disease neuroimaging initiative-based evaluation study. Journal of
  Alzheimer's Disease 71:165--183.
\bibAnnoteFile{tustison2019longitudinal}

\end{thebibliography}

\end{document}